\documentclass[useAMS,usenatbib,article] {mn2e}
\usepackage{epsfig}
\usepackage{graphicx}
\usepackage{amssymb} 

\newcommand{\ratio}{\mbox{$v_{\infty}$/$v_{\rm esc}$}}

\title[SFE and age spread in massive clusters]{Feedback-regulated star formation: II.  dual constraints on the SFE and the age spread of stars in massive clusters}

\author[S. Dib et al.]{Sami Dib$^{1,2,3}$\thanks{E-mail: sami.dib@gmail.com}, Julia Gutkin$^{4}$, Wolfgang Brandner$^{3}$,  Shantanu Basu$^{5}$\\ 
$^{1}$Astronomisches Rechen-Institut, Zentrum f\"{u}r Astronomie der Universit\"{a}t Heidelberg, M\"{o}nchhofstr. 12-14, 69120, Heidelberg, Germany\\
$^{2}$School of Astronomy, Institute for Research in Fundamental Sciences (IPM), PO Box 19395-5531, Tehran, Iran\\
$^{3}$Max-Planck Institute f\"{u}r Astronomie, K\"{o}nigstuhl 17, 69117, Heidelberg, Germany\\ 
$^{4}$UPMC-CNRS, UMR7095, Institut d'Astrophysique de Paris, F-75014, Paris, France\\
$^{5}$Department of Physics and Astronomy, University of Western Ontario, London, Ontario, N6A 3K7, Canada}

\begin{document}
\maketitle

\date{Accepted XXX. Received XXX}

\pagerange{\pageref{firstpage}--\pageref{lastpage}}
\pubyear{2009}
\label{firstpage}

\begin{abstract} 

We show that the termination of the star formation process by winds from massive stars in protocluster forming clumps imposes dual constraints on the star formation efficiencies (SFEs) and stellar age spreads ($\Delta \tau_{*}$) in stellar clusters. We have considered two main classes of clump models. One class of models in one in which the core formation efficiency (CFE) per unit time and as a consequence the star formation rate (SFR) is constant in time and another class of models in which the CFE per unit time, and as a consequence the SFR, increases with time. Models with an increasing mode of star formation yield shorter age spreads (a few $0.1$ Myrs) and typically higher SFEs than models in which star formation is uniform in time. We find that the former models reproduce remarkably well the SFE$-\Delta \tau_{*}$ values of starburst clusters such as NGC 3603 YC and Westerlund 1, while the latter describe better the star formation process in lower density environments such as in the Orion Nebula Cluster. We also show that the SFE and $\Delta \tau_{*}$ of massive clusters are expected to be higher in low metallicity environments. This could be tested with future large extragalactic surveys of stellar clusters. We advocate that placing a stellar cluster on the SFE-$\Delta \tau_{*}$ diagram is a powerful method to distinguish between different stellar clusters formation scenarios such as between generic gravitational instability of a gas cloud/clump or as the result of cloud-cloud collisions. It is also a very useful tool for testing star formation theories and numerical models versus the observations. 
            
\end{abstract} 


\section{INTRODUCTION}\label{intro}

The photometric, structural, and dynamical properties of young clusters yield a number of constraints for star formation theories. Two of the most important constraints are the star formation efficiency (SFE) and the spread of stellar ages, $\Delta \tau_{*}$. The SFE in a protocluster region is defined as being the mass fraction of gas which is converted into stars at the end of the process of star formation. Theoretically, the  SFE can be defined as SFE $\approx {M_{cluster}}/{M_{clump}}$, where $M_{cluster}$ is the mass of the stellar cluster and $M_{clump}$ is the initial mass of the gas clump to which can be added the mass of gas accreted by the clump from its environment before the entire star formation process ends. However, in the observations, it is impossible to estimate what the original clump mass was, once a large fraction of the gas has been expelled from the protocluster region. In embedded and semi-embedded clusters, the SFE is usually defined as being $SFE \approx  M_{cluster}/{(M_{cluster}+M_{gas})}$, where in this case $M_{gas}$ is the remaining mass of gas in the cluster. The SFEs derived from observations of embedded clusters using this equation fall in the range $\approx [0.1-0.6]$ (Lada \& Lada 2003). The SFEs found in protocluster forming clumps tend to be generally larger than those measured on the scale of entire Giant Molecular Clouds (GMCs) or in regions of low clustered star formation. Myers et al. (1986), Evans et al. (2009), and Lada et al. (2010) report values of the SFE in GMCs that fall in the range of $\approx [0.01-0.1]$. Federrath \& Klessen (2013) found a lower limit of the SFE in the Serpens cloud of $\gtrsim 0.08$.  

As stars form in the densest regions of molecular clumps/clouds (i.e., in dense, gravitationally bound cores), it is crucial to assess what is the core formation efficiency (CFE) per unit time in the protocluster forming clumps and what are its effects on the final SFE. Supersonic turbulence that is ubiquitously observed in molecular clouds produces local compressions of which a fraction can be 'captured' by gravity and proceed to collapse into stars (e.g., Klessen 2000; Dib et al. 2007a,2008a; Dib \& Kim 2007). Magnetic fields play also an important role in determining the mass fraction of gravitationally bound gas. Results from both ideal and non-ideal MHD simulations show that stronger magnetic fields (in terms of magnetic criticality) lower the rate of dense core formation in the clouds (e.g., V\'{a}zquez-Semadeni et al. 2005; Price \& Bate 2008; Basu et al. 2009; Dib et al. 2010a;  Collins et al. 2012; Federrath \& Klessen 2012). Dib et al. (2010a) showed that the CFE per unit free-fall time of the cloud, CFE$_{ff}$, are of the order of $\sim 6~\%$ and $\sim 33~\%$ for clouds with mass to-magnetic flux ratios of $\mu=2.2$ and $8.8$, respectively (with $\mu$ being normalised by the critical value for collapse).

In spite of the role played by magnetic fields and turbulence, the instantaneous SFE will continue to increase in bound clumps/clouds until either the gas reservoir is exhausted or otherwise when gas is expelled from the cloud. The role of stellar feedback in setting the final value of the SFE by totally removing the gas from the cloud has been studied by several authors. The role of protostellar outflows has been investigated numerically by Banerjee et al. (2007), Li \& Nakamura (2006), Wang et al. (2010), and Hansen et al. (2012). While protostellar outflows may play a role in generating a self-sustained turbulence in a protocluster forming region, they do not seem to inject enough energy that can lead to the gas removal from the region (e.g., Nakamura \& Li 2007). Another form of stellar feedback is associated with O and B stars, in their main sequence phase, and beyond. OB stars emit ultraviolet radiation which ionises the surrounding gas and heats it to temperatures of $\sim 10^{4}$ K. This warm and ionised bubble provides the environment in which particles accelerated from the stellar surface by interaction with a fraction of the stellar radiation propagate outwards. The effect of ionising radiation from massive stars in the evolution of clouds and their SFR has been considered in the time-dependent, semi-analytical models of Goldbaum et al. (2011), and Zamora-Avil\'{e}s et al. (2012)

The age spread of stars ($\Delta \tau_{*}$) is another important signature that can help characterise the mode of star formation in protocluster forming regions (i.e., uniform in time, accelerated, or eventually sequential) and can inform on the relevance of stellar feedback and its associated timescales for the process of gas expulsion from the clump and the quenching of star formation. The work of Palla \& Stahler (1999,2000) on a sample of nearby Galactic clusters suggested that the age spreads of stars in the clusters could be as large as $\sim 10$ Myrs with the bulk of star formation occurring at advanced times (also Herbst \& Miller 1982; see however Stahler 1985 for counter arguments). This was interpreted by Palla \& Stahler as being the result of the slow buildup of the cloud in the initial phase which is followed by a stage of gravitational contraction. However, these large age spreads have been contested by Hartmann (2001;2003) who argued that both theoretical uncertainties in the Pre-Main Sequence (PMS) stars evolutionary tracks as well as observational uncertainties preclude a robust determination of the true age spreads. Pflamm-Altenburg \& Kroupa (2007) showed that larger age spreads can be produced in young stellar clusters if field stars fall into the newly generated deeper potential of the contracting protocluster cloud. Nevertheless, Hartmann et al. (2012)  have recognised that some low levels of star formation are to be expected during the early evolutionary stages of star-forming clouds. 

Preibisch (2012) considered a statistically significant sample of $10^{5}$ coeval age stars in a cluster and placed them of the HR diagram after assigning them with observational uncertainties typical of the Upper Sco cluster ($\approx 0.15$ magnitudes due to the effects of photometric variability, unresolved binaries, and spreads in the distances of stars within the cluster). The inferred age distribution he obtained by constraining the noised data by theoretical isochrones indicates, wrongly, an age spread of $\approx 10$ Myrs, with an age distribution which is similar to the ones obtained by Palla \& Stahler. A cautionary note about the misinterpretation of observed spreads in the luminosity of stars as an indication of the existence of a true age spread has also been made by Hillenbrand et al. (2008). Jeffries et al. (2011) revisited the age distribution of stars in the Orion Nebula Cluster. They argued that the duration of the star formation process, and hence the age spread, should not be larger than the median age of protostellar discs which is a few Myrs at most, and possibly shorter. In the low efficiency and low-mass forming region of Taurus, Kraus \& Hillenbrand (2009) measured age spreads of $\gtrsim 2$ Myrs. Age spreads of $\gtrsim 2$ Myrs were measured by Bik et al. (2012) in the OB association W3 and similar results were obtained by Wang et al. (2011) in the massive star forming region S255, while Ojha et al. (2011) suggested that the age spread in S255 could be as large as 4 Myrs. An age spread in the range of $2.8-4.4$ Myrs has been inferred for the stellar association LH 95 in the Large Magellanic Cloud by Da Rio et al. (2010a). In contrast, in massive and high stellar surface density clusters (usually referred to as starburst clusters), the measured values of $\Delta \tau_{*}$ are found to be $\lesssim 1$ Myrs. Recently, Kudryavtseva et al. (2012) used Bayesian statistics on the photometric properties of the starburst clusters Westerlund 1 and NGC 3603, coupled to an assessment of cluster membership using astrometry and inferred age spreads of 0.4 and 0.1 Myrs for these two clusters, respectively (see also Clark et al. 2005).  

In this work, we investigate the constraints from stellar feedback on both the SFE and the age spread of stars in clusters as a function of the protocluster clumps mass and for various modes of star formation (i.e., uniform in time vs. accelerated). Models based on the regulation of the star formation process by feedback from massive stars have been very successful in reproducing several observations such as the Kennicutt-Schmidt relations both on cloud scales and galactic scales (Dib 2011a,b; Zamora-Avil\'{e}s et al. 2012), the dependence of the star formation efficiencies on the metallicity (Dib et al. 2011), and the metal enrichment of the host galaxy and the clusters self-enrichement (e.g., W\"{u}nsch et al. 2008). In \S.~\ref{model}, we describe the protocluster clump model, the dense core model, the distributions of cores that form in the clump (i.e., the initial core mass function), and the adopted feedback model from radiation driven stellar winds. In \S.~\ref{results} we present the co-evolution of the prestellar core mass function and of the IMF in detail for a fiducial case with solar metallicity and with a constant value of the CFE per unit time. We then compare the results of models with various core/star formation histories and various clump masses. Finally, we show that the regulation of star formation by massive stars in protocluster forming clumps yields values of the SFE and age spreads which match perfectly the available observational data for massive clusters. In \S.~\ref{summary}, we summarise our results.       

\begin{table}
\centering
\begin{minipage}{7.5cm}
\begin{tabular}{l l }
\hline
Clump variables & Meaning of the variables \\
\hline
$M_c$ & mass\footnote{The fiducial value is $10^{5}$ M$_{\odot}$ and $M_{c}$ is varied in the range $[5\times10^{4}-5\times10^{5}]$ M$_{\odot}$} \\
$R_{c}$ &  radius \\
$R_{c0}$ & core radius \\
$v_{c}$ & velocity dispersion of the gas \\
$\rho_{c0}$ & central density  \\
$E_{grav}$ & gravitational energy \\
$\alpha$ & exponent of the $v_{c}-R_{c}$ relation\\
$t_{ff,c}$ & free-fall timescale \\ 
$\gamma$ & PDF of density field parameter\footnote{We adopt a fiducial value of $\gamma=0.26$. We also explore the effect of different $\gamma$ values in the range [0.26-0.65]}\\
$CFE(t)$~~~~ &  core formation efficiency per unit time \\
\hline
Core variables & Meaning of the variables \\
\hline
$M$ & mass  \\
$R_{p}$ & radius  \\
$R_{p0}$ & core radius   \\
$\rho_{p0}$ & central density  \\
$t_{ff}$ & free-fall timescale  \\
$t_{cont,p}$ & contraction timescale \\
$\mu$ & exponent of the $\rho_{p0}-M$ relation\footnote{We adopt a value of $\mu=0.2$. Observations indicate a range of [0-0.6] (Caselli \& Myers 1995; Johnstone \& Bally 2006)} \\
$\nu$ &  $\nu=(t_{cont,p}/t_{ff})$\footnote{We adopt a value of $\nu=3$. Observations and theoretical considerations indicate a range of [1-10] (e.g., McKee 1989)} \\
\hline
Feedback variables & Meaning of the variables\\ 
\hline
$E_{wind}$ & time integrated energy from massive stars \\
$E_{k,wind}$ &  $E_{k,wind}=\kappa E_{wind}$\\
$\phi$ & core-to-star efficiency parameter\footnote{We adopt a value of $\phi=1/3$. Theoretical models suggest a value in the range of [0.3-0.7] (Matzner \& Mckee 2000)}\\
$\kappa$ & wind efficiency parameter\footnote{We vary $\kappa$ in the range of [0.01-0.2]} \\

\hline
 \end{tabular}
\end{minipage}
\caption{Main variables and parameters in the model. The top panel describes the protocluster clump variables/parameters, the middle panel the dense cores variables/parameters, and the bottom panel the stellar feedback variables/parameters.}
\label{tab1}
\end{table}

\section{DESCRIPTION OF THE MODEL}\label{model}

The model used in this work follows star formation in a protocluster forming clump and the effect of stellar feedback in expelling the gas from the clump and setting the final SFE and age spread of stars in the newborn cluster. The model describes the co-evolution of the populations of gravitationally bound cores and of the IMF that form in the clump. The assumption is made that populations of gravitationally bound cores form, with a prescribed core formation efficiency per unit time, at different epochs and at different positions in the protocluster clump as a consequence of its gravo-turbulent fragmentation. The distribution of masses for each generation of gravitationally bound cores that form in the cloud is prescribed by a gravo-turbulent fragmentation model and depends on the local structural and dynamical conditions (i.e., the local Mach number, number density, and the power spectrum of turbulent velocities at each radial position from the clump centre). The model is one-dimensional and assumes spherical symmetry. The effects of gravo-turbulent fragmentation of the clump are treated in a phenomenological way and are based on previous theoretical and numerical works (Eqs. 6-10 below). The cores that form in the clump have finite lifetimes which are taken to be a few times their free-fall timescales and which depend on their masses and positions in the clump, after which they collapse to form stars and gradually populate the local IMF. Stellar winds from the newly formed massive stars (M$_{*} > 5$ M$_{\odot}$) inject kinetic energy in  the clump and a fraction of the wind energy will effectively cause the expulsion of the gas from the protocluster region. This halts the formation of newer generations of cores and quenches any subsequent star formation, thus setting the final SFE and age spread of stars in the newly formed cluster. In the following sections, we describe in detail the different components of the model. Tab.~\ref{tab1} lists the main variables and parameters of the model, their meaning, and when appropriate, the fiducial value of the parameters and their ranges.  

\subsection{PROTOCLUSTER CLUMPS}\label{clusters}

Several studies have established that star clusters form in dense ($\gtrsim 10^{3}$ cm$^{-3}$) clumps embedded in a lower density parental molecular cloud (e.g., Lada \& Lada 2003; Csengeri et al. 2011). Using  the C$^{18}$O $J=1-0$ molecular emission line, Saito et al. (2007) presented a study for a large sample of cluster forming clumps whose masses and radii vary between [15-1500] M$_{\odot}$Êand [0.14-0.61] pc, respectively. Dib et al. (2010b) used the data of Saito et al. (2007) and derived the mass-size and velocity dispersion-size relations of the clumps which are given by $M_{clump}({\rm M_{\odot}})=10^{3.62 \pm 0.14} R_{c}^{2.54 \pm 0.25} ({\rm pc})$ and $v_{c}({\rm km~s^{-1}})=10^{0.45 \pm 0.08} R_{c}^{0.44 \pm 0.14}({\rm pc})$, where $R_{c}$ is the size of the clump. In the study of Mueller et al. (2002), the average slope of the density profiles of 51 star forming clumps is $-1.8 \pm 0.4$. In this work, we assume that protocluster clumps are spherically symmetric
and adopt a protocluster clump model that follows an $r^{-2}$ density profile:

\begin{equation} 
\rho_{c}(r)= \frac{\rho_{c0}} {1+(r/R_{c0})^{2}},
\label{eq1}
\end{equation}

\noindent where $r$ is the distance from the clump centre, $R_{c0}$  is the clump's core radius (core radius here stands for the central region of the clump), and $\rho_{c0}$ is the density at the centre. For a given mass of the clump,  $M_{clump}$, the central density $\rho_{c0}$ is given by:

\begin{equation} 
\rho_{c0}= \frac{M_{clump}}{4 \pi R_{c0}^{3} [(R_{c}/R_{c0})-\arctan(R_{c}/R_{c0})]}.
\label{eq2}
\end{equation}

The temperatures of the cluster forming clumps are observed to vary between 15 and 70 K (e.g., Saito et al. 2007). We assume that the equation of state in the clumps is isothermal with $T=20$ K. We also adopt a value for the core radius of the clump of $R_{c0}=0.2$ pc, which is very similar to the core radius value of young massive clusters such as the Orion Nebula Cluster (Hillenbrand \& Hartmann 1998). We relate the sizes of the modelled protocluster clumps to their masses using these relations. We also assume that the velocity dispersion-size relation derived on clumps scales by Saito et al. (2007) remains valid on scales smaller that the clumps outer radii, i.e., $\sigma_{v}(r)=10^{0.45\pm0.08} r^{0.44\pm0.14}$, where $\sigma_{v}(r)$ is the velocity dispersion of the gas at a distance $r$ from the clump's centre. The clump is divided into a number of radial shells (100 bins), in which a local velocity dispersion of the gas and thus a Mach number can be calculated. As turbulent fragmentation is efficient in compressing the gas when turbulent motions are supersonic, we require that $\sigma_{v}(r) \geq c_{s}$, where $c_{s}$ is the speed of sound at the adopted temperature of $T=20$ K. 

\subsection{PROTOSTELLAR CORES}\label{core_model}

As discussed above, a fraction of the mass of the clump is converted into gravitationally bound cores with a given global core formation efficiency per unit time. In this model, gravitationally bound cores are over-dense regions with respect to the local density profile of the clump at the position they are formed. In the following, we describe the protostellar core model. Whitworth \& Ward-Thompson (2001) applied a family of Plummer sphere-like models to the contracting dense core L1554, which is representative of the population of gravitationally bound cores that can be found in the clumps that are considered in this work. They found a good agreement with the observations if the density profile of the core has the following form $\rho_{p}(r_{p})= {\rho_{p0}}/[1+(r_{p}/R_{p0})^{2}]^{2}$, \noindent where $\rho_{p0}$ and $R_{p0}$ are the central density and core radius of the core, respectively. We require that the size of the core, $R_{p}$, depend on both its mass, $M$, and position within the clump, $r$. The dependence of $R_{p}$ on $r$ requires that the density at the edges of the core equals the ambient clump density, i.e., $\rho_{p}(R_{p})=\rho_{c}(r)$ (i.e., pressure balance at constant temperatures). This would result in smaller radii for cores of a given mass when they are located in the inner parts of the clump.

 We assume that the cores contract on a timescale, $t_{cont,p}$ which we take to be a few times their free fall timescale $t_{ff}$, and which is parametrized by $t_{cont,p}(r,M)= \nu ~ t_{ff}(r,M)= \nu \left( 3 \pi/32~G \bar{\rho_{p}} (r,M) \right)^{1/2}$, where $G$ is the gravitational constant, $\nu$ is a constant $\ge 1$ and $\bar{\rho_{p}}$ is the radially averaged density of the core of mass $M$, located at position $r$ in the clump. Theoretical considerations suggest that $t_{cont,p}$ can vary between $t_{ff}$ ($\nu=1$) and $10~t_{ff}$ ($\nu=10$) with the latter value being the characteristic timescale of ambipolar diffusion (McKee 1989; Fiedler \& Mouschovias 1992; Ciolek \& Basu 2001). Both observational and numerical estimates of gravitationally bound cores lifetimes tend to show that they are of the order of a few times their free-fall time (e.g., Galv\'{a}n-Madrid et al. 2007). In all models, we adopt a value of $\nu=3$. We assume that the density contrast between the centre and the edge of the cores depends on their masses following a relation of the type $\rho_{p0}  \propto M^{\mu}$. The density contrast between the centre and the edge for a core with the minimum mass we are considering, $M_{min}$ (typically $M_{min}=0.1$ M$_{\odot}$) is 15 (slightly higher than the value of 14.1 for a stable Bonnor-Ebert sphere). Thus, for cores with masses $M > M_{min}$, the density contrast between the centre and the edge of the core will be equal to $15\times (M/M_{min})^{\mu}$. Available observations show that  $\mu$ falls in the range of $\approx [0-0.6]$ (Caselli \& Myers 1995; Johnstone \& Bally 2006). In this work, we adopt an intermediate value of 0.2.
 
Once a core's age is equal to $t_{cont,p}$, the core is collapsed to form a single star (i.e., with no sub-fragmentation). We assume that only a fraction of the mass of the core ends up locked in the star. We account for this mass loss in a purely phenomenological way by assuming that the mass of a star which is formed out of a core of mass $M$ is given by $M_{*}=\phi M$, where $\phi \leq 1$. Matzner \& McKee (2000) showed that $\phi$ can vary in the range $0.25-0.7$. A comparison of HD and MHD simulations of dense core formation with observational data by Heiderman et al. (2010) shows that the core-to-star efficiency can range from $0.3$ to $0.7$ (Federrath \& Klessen 2012). The value of $\phi \approx 1/3$ is commonly quoted and used in the literature. This is motivated by the similarity between the core mass function and the Kroupa/Chabrier IMFs found by Alves et al. (2007) and by Andr\'{e} et al. (2010) in the Pipe dark cloud and in the Aquila star forming region, respectively. In this work, we adopt a value of $\phi=1/3$ and assume that it is independent of the mass of the cores.
 
\subsection{FEEDBACK}\label{feedback}

We assume that the formation of cores in the protocluster clump, and consequently star formation, is terminated whenever the fraction of the wind energy stored into motions that oppose gravity exceeds the gravitational binding energy of the clump. Whenever this occurs (at $t=t_{exp}$), the gas is expelled from the protocluster clump and star formation is quenched. Thus, at any epoch $t < t_{exp}$, gas is removed from the clump only to be turned into stars. We take into account the feedback generated by the stellar winds of massive stars ($M_{\star} \ge 5$ M$_{\odot}$). In order to quantify the power of stellar winds, we used of a modified version of the stellar evolution code CESAM (Piau et al. 2011) to calculate a grid of main sequence stellar models for stars in the mass range [5-80] M$_{\odot}$ (with steps of 5 M$_{\odot}$) at various metallicities between $Z/Z_{\odot}=1/10$ and $Z/Z_{\odot}=2$. The evolution of massive stars was followed using the CESAM code for 1 Myr on the main sequence. In Dib et al. (2011), we showed that the time variations of quantities such as the stellar luminosity, effective temperature, and stellar radius, are of the order of a few percent, increasing with time to about 10-20 percent at most for the most massive stars. Since we are interested in the early time evolution of protoclusters, we average the stellar properties over the first $5 \times 10^{5}$ yr and use the averaged stellar properties as characteristic values for stars on the main sequence. This procedure has been repeated for all metallicities. In a second step, we use the grid of calculated time averaged stellar properties to evaluate, for the different metallicity cases, the stellar mass loss rates and the power of the stellar winds. To that purpose, we use the results of the stellar atmosphere models developed by Vink et al. (2001). These models allow for the evaluation of the stellar mass loss rate $\dot{M}_{\star}$, as a function of the stellar mass $M_{\star}$, effective temperature $T_{eff}$, the stellar luminosity $L_{\star}$, the metallicity $Z$, and the ratio of the velocity of the wind at infinity to the escape velocity, $\ratio$. Vink et al. (2001) did not derive the values of $v_{\infty}$, therefore, we use instead the relations obtained by Leitherer et al. (1992). In Dib et al. (2011) we have calculated the rate of mechanical energy deposited by the winds, $\dot{E}_{wind}(M_{\star})=\dot{M}_{\star}~v_{\infty}^{2}$, and provided polynomial fits of $\dot{E}_{wind}$ as a function of $M_{\star}$ for the various metallicity cases (i.e., fourth order polynomial fits in $M_{*}$). We refer the reader to Dib et al. (2011) for additional details on the derivation of $\dot{E}_{wind}$. The total kinetic energy from the winds, at any given epoch, and position r in the clump, following the formation of massive stars in the clump, is given by:

\begin{equation}
E_{wind}(r) = \int_{t'=0}^{t'=t} \int_{M_{\star}=5~\rm{M_{\odot}}}^{M_{\star}=120~\rm{M_{\odot}}} \left( \frac{N(M_{\star},r) \dot{E}_{wind}(M_{\star})}{2} dM_{\star}\right) dt',
\label{eq3}
\end{equation}

where $N(M_{\star},r)$ is the number of stars of mass $M_{\star}$ that are present at a distance $r$ from the centre of the clump, at time $t$. The total wind energy $E_{wind}$ is calculated as being the sum of all the radial values $E_{wind}=\Sigma E_{wind} (r)$. We assume that only a fraction of $E_{wind}$ will be transformed into systemic motions that can oppose gravity and cause the evacuation of the gas from the protocluster clump. The rest of the energy is assumed to be dissipated in wind-wind collisions or escape the wind bubble. The effective kinetic wind energy is thus given by: 

\begin{equation}
 E_{k,wind}=\kappa~E_{wind},
\label{eq4}
\end{equation}
 
\noindent where $\kappa$ is a free parameter that is $\leq 1$. The exact value of $\kappa$ may well vary from system to system depending on the number of massive stars, their locations within the cluster, and the details of their wind interactions. Given that $\kappa$ is not constrained by any current observations or theoretical model, we vary $\kappa$ over a wide range of [0.01, 0.2]. $E_{k,wind}$ is compared at every time-step to the value of the gravitational energy of the clump, $E_{grav}$, which is calculated as being:

\begin{equation}
E_{grav} =  -\frac{16}{3} \pi^{2} G \int_{0}^{R_{c}} \rho_{c}(r)^{2} r^{4}  dr,
\label{eq5}
\end{equation}

\noindent where $\rho_{c}$ is given by Eq.~\ref{eq1}. Gas is expelled from the cluster and star formation is terminated when the ratio $E_{k,wind}/E_{grav}$ reaches unity for a given value of $\kappa$.

\subsection{INITIAL CORE MASS FUNCTIONS} \label{cmf}

\begin{figure*}
\begin{center}
\includegraphics[height=18.5cm, width=15.5cm]{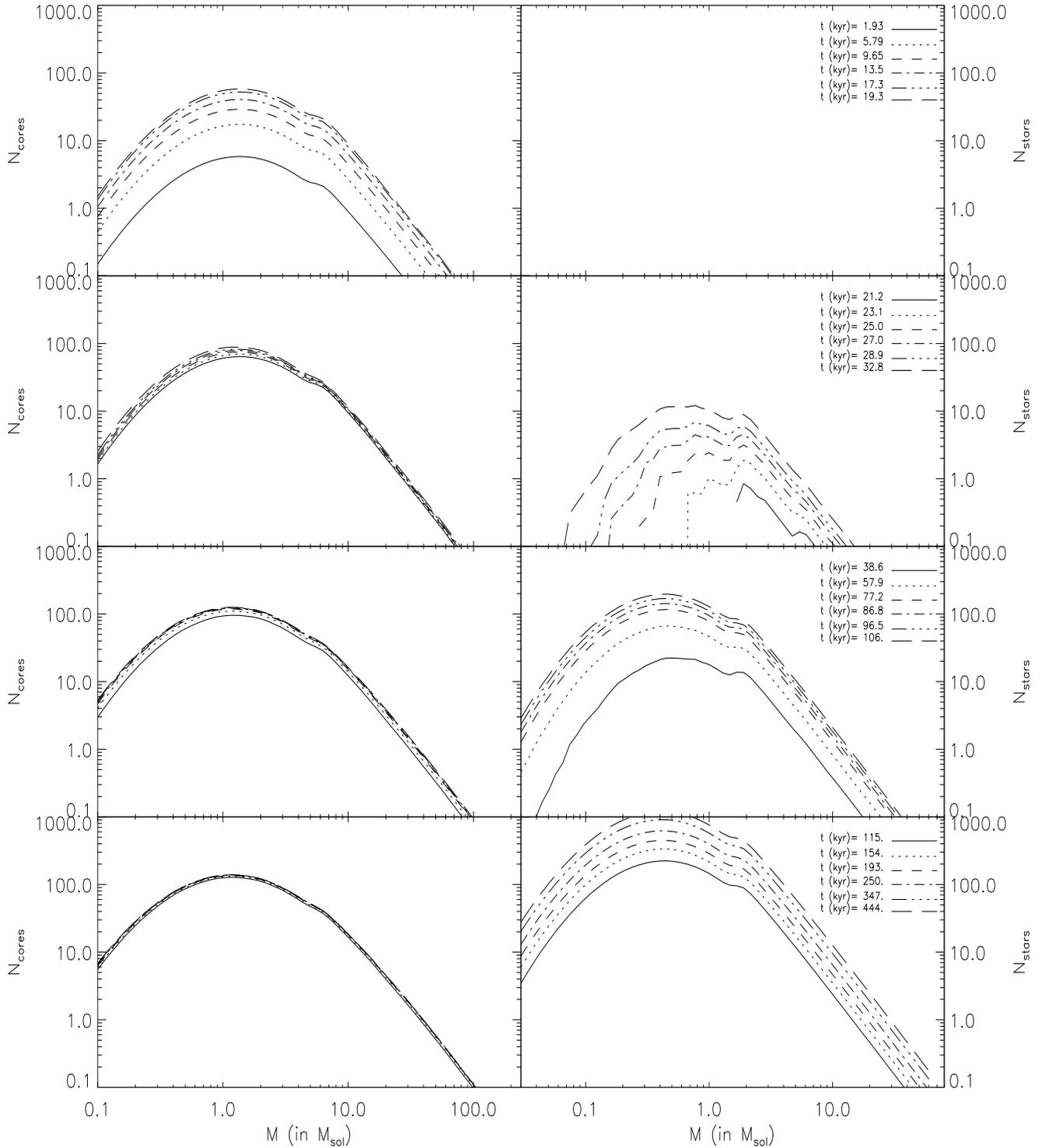}
\end{center}
\vspace{0.7cm}
\caption{Time evolution of the pre-stellar core mass function (left), and stellar mass function (right) in the protocluster clump with the fiducial model parameters. The last  time-step shown is at $t=4.4\times 10^{5}$ yrs and corresponds to the epoch at which gas is expelled from the protocluster clump for a value wind efficiency parameter $\kappa=0.01$. The mass of the protocluster clump in this model is $10^{5}$ M$_{\odot}$ and the metallicity is $Z=Z_{\odot}$.}
\label{fig1}
\end{figure*}

As stated above, the formation of dense cores in the clump is assumed to be the result of its gravo-turbulent fragmentation. As in our previous work (Dib et al. 2010b,2011), we use the formulation given by  Padoan \& Nordlund (2002) in order to calculate the local distributions of gravitationally bound cores that form at different radii in the clump. In the following, we briefly remind what the constituents of this model are. The model assumes that the probability distribution function of the density field of  an isothermal, turbulent, compressible gas is well described by a lognormal distribution (V\'{a}zquez-Semadeni 1994) and is given by:

\begin{equation}
P(\ln x) d\ln~x=\frac{1}{\sqrt{2 \pi \sigma_{d}^{2}}} \exp \left[-\frac{1}{2} \left(\frac{\ln x-{\bar{\ln x}}}{\sigma_{d}} \right)^{2} \right] d\ln~x,
\label{eq6}
\end{equation}
  
\noindent where $x$ is the number density normalised by the average number density, $x=n/\bar{n}$. The standard deviation of the density distribution $\sigma_{d}$ and the mean value $\bar {\ln x}$ are functions of the local thermal rms Mach number, $\cal M$, and  $\bar{\ln x}=-\sigma^{2}_{d}/2$ with $\sigma^{2}_{d}=\ln(1+{\cal M}^{2} \gamma^{2})$. Padoan \& Nordlund (2002) suggested a value of $\gamma \approx 0.5$, whereas Kritsuk et al. (2007) using higher resolution simulations found that $\gamma \approx 0.260 \pm 0.001$ The latter value is the one adopted in our models. Federrath et al. (2008;2010) showed that the value of $\gamma$ may depend on the nature of the turbulence driver. They found values of $\gamma \approx 0.33$ when turbulence is driven by purely solenoidal motions and up to $\gamma \approx 1$ when turbulence is driven by purely compressive motions. A second step in this approach is to determine the mass distribution of dense cores. Padoan \& Nordlund (2002) showed that by making the following assumptions: (a) the power spectrum of turbulence is a power law and, (b) the typical size of a dense core scales as the thickness of the post-shock gas layer, the core mass spectrum is given by:

\begin{equation}
N(M)~d\log~M \propto M^{-3/(4-\theta)} d\log~M,
\label{eq7}
\end{equation}   
  
\noindent where $\theta$ is the exponent of the kinetic energy power spectrum, $E_{k} \propto k^{-\theta}$, and is related to the exponent $\lambda$ of the size-velocity dispersion relation in the cloud with $\theta=2 \lambda+1$ (in this work $\lambda=0.44$, see \S.~\ref{clusters}). However, Eq.~\ref{eq7} can not be directly used to estimate the number of cores that are prone to star formation. It must be multiplied by the local distribution of Jeans masses. At constant temperature, this distribution becomes:
 
\begin{eqnarray}
N (r,M)~d\log~M =f_{0}(r)~M^{-3/(4-\theta)} \nonumber \\
            \times \left[\int^{M}_{0} P(M_{J}) dM_{J}\right]d\log~M,
\label{eq8}
\end{eqnarray}   

The local normalisation coefficient $f_{0}(r)$ is obtained by requiring that $\int^{M_{max}}_{M_{min}} f_{0} (r) N (r,M)~dM=1$ in a shell of width $dr$, located at distance $r$ from the clump's centre. The local distribution of Jeans masses, $P(M_{J})$, is given by (Padoan \& Nordlund 2002):

\begin{equation}
P(M_{J})~dM_{J}=\frac{2~M_{J0}^{2}}{\sqrt{2 \pi \sigma^{2}_{d}}} M^{-3}_{J} \exp \left[-\frac{1}{2} \left(\frac{\ln~M_{J}-A}{\sigma_{d}} \right)^{2} \right] dM_{J},
\label{eq9}
\end{equation}   

\noindent where $M_{J0}$ is the Jeans mass at the mean local density. Therefore, the local distribution of cores generated in the clump, at an epoch $\tau$, $N(r,M,\tau)$ would be given by:

\begin{equation}
{N} (r,M,\tau) dt=\frac{{\rm CFE} (r,t) \rho_{c}(r)} {<M>(r)~t_{cont,p} (r,M)} \frac{dt}{t} N(r,M),
\label{eq10}
\end{equation} 

where $dt$ is the time interval between two consecutive epochs, and CFE $(r,t)$ is a parameter smaller than unity which describes the local mass fraction of gas that is transformed into cores per unit time. In the present study, we assume that the CFE is independent of $r$ so that CFE$(r,t)$=CFE$(t)$. If the core formation process is constant in time, then the term CFE $(t) (dt/t)$ can be taken as CFE$_{ff} (dt/t_{ff,cl})$ where CFE$_{ff}$ is a constant that describes the core formation efficiency per unit free-fall time of the clump $t_{ff,cl}=\sqrt{(3\pi/32 G {\bar \rho}_{c})}$ and where $\bar{\rho}_{c}$ is the average density in the clump. In Eq.~\ref{eq10}, the term CFE $(t)\rho_{c}(r)/t_{cont,p (r,M)})$ is the local rate of formation of dense cores in units of mass per unit volume and per unit time. When divided by the average local core mass $<M> (r)$ (for the local CMF between the minimum and maximum cores masses taken here to be $0.1$ M$_{\odot}$ and 210 M$_{\odot}$, respectively), it gives the number of cores formed per unit volume and unit time. This is then multiplied by the term ($dt/t$) to account for the total number of cores that form in the time interval $dt$.   

\section{RESULTS}\label{results}

In the sections below, we present the results for an ensemble of models. We explore cases in which the CFE is uniform in time (i.e., that have a constant CFE$_{ff}$, and call them C-class models), as well as cases where the CFE increases as a function of time (PL and EXP-class models when the increase of the CFE is described by a power law function or an exponential function, respectively). The latter models are intended to mimic a global gravitational contraction of the clump and an increase in the gravitational boundedness of a larger fraction of the clumps's mass.  In our models, the SFE is defined as being: 

\begin{equation}
{\rm SFE}=\frac{M_{cluster}}{M_{clump}},
\label{eq11}
\end{equation}

\noindent where $M_{cluster}$ is the final mass of the cluster (i.e., after gas has been expelled from the clump), and $M_{clump}$ is the initial clump mass. The age spread of stars in the cluster is defined as being:

\begin{equation}
\Delta\tau_{*}=t_{exp}-t(N_{*}=1),
\label{eq12}
\end{equation}

\noindent where $t_{exp}$ is the epoch at which the gas expulsion occurs and $t(N_{*}=1)$ is the epoch at which the number of stars, in any stellar mass bin is equal to unity. In principle, the local CFE per unit time (i.e., at different radial positions from the clumps centre) can be linked to the local value of the virial parameter, Mach number, magnetic field strength, and to the nature of the turbulence driver ; i.e., a turbulence driver that generates predominantly shearing or compressive motions (see Federrath \& Klessen 2012). However, in our models, we do not explicitly account for the value of the magnetic field nor for the nature of the turbulence driver. We therefore treat the CFE per unit time as a free parameter whose value (or normalisation in the case of the PL- and EXP-class models) is bracketed by the ranges found in numerical simulations (e.g., Dib et al. 2010; Federrath \& Klessen 2012). 

\subsection{A FIDUCIAL CASE AND INFLUENCE OF THE CORE FORMATION HISTORY}\label{fiducial}

\begin{figure}
\begin{center}
\epsfig{figure=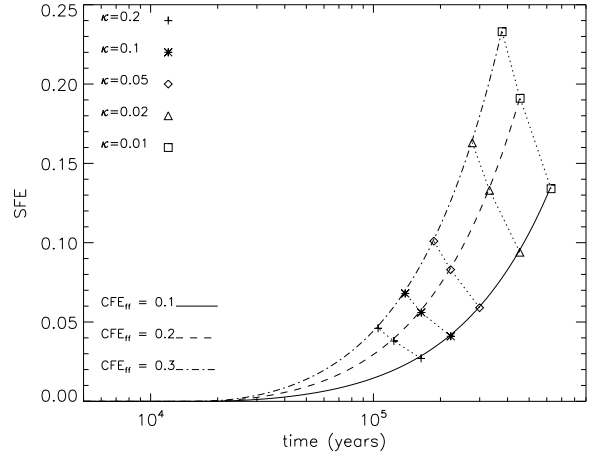,width=\columnwidth} 
\end{center}
\caption{time evolution of the SFE in three models with a constant core formation efficiency per unit time. The different symbols mark the value of the SFE and the expulsion epoch of the gas $t_{exp}$ for various values of the wind efficiency parameter, $\kappa$. The mass of the clump in all of these models is $10^{5}$ M$_{\odot}$ and the metallicity is $Z=Z_{\odot}$.}
\label{fig2}
\end{figure}

\begin{figure}
\begin{center}
\epsfig{figure=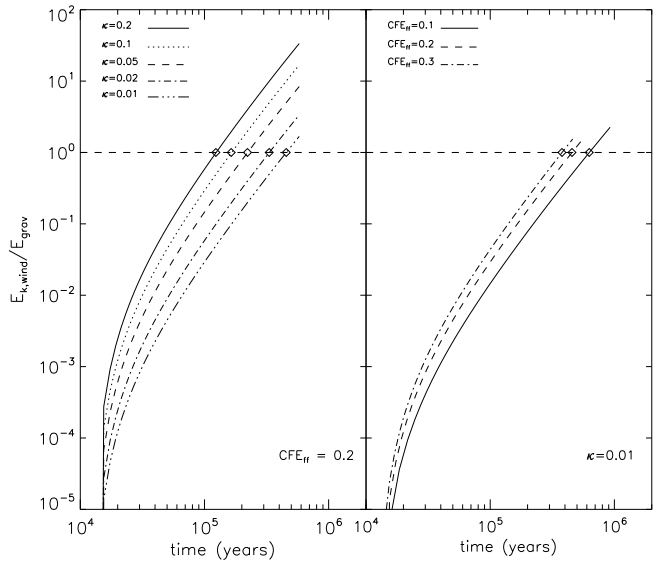,width=0.9\columnwidth} 
\end{center}
\vspace{0.5cm}
\caption{time evolution of the ratio of effective wind energy to the gravitational energy of the clump in a) (right panel) three models with a constant core formation efficiency per unit time and for a given value of $\kappa=0.01$ and b) (left panel) all with CFE$_{ff}=0.2$ but assuming different values of the wind efficiency parameter, $\kappa$. The diamonds mark the  epoch at which gas is expelled from the cluster, $t_{exp}$. The mass of the clump in all of these models is $10^{5}$ M$_{\odot}$ and the metallicity is $Z=Z_{\odot}$.}
\label{fig3}
\end{figure}

\begin{figure}
\begin{center}
\epsfig{figure=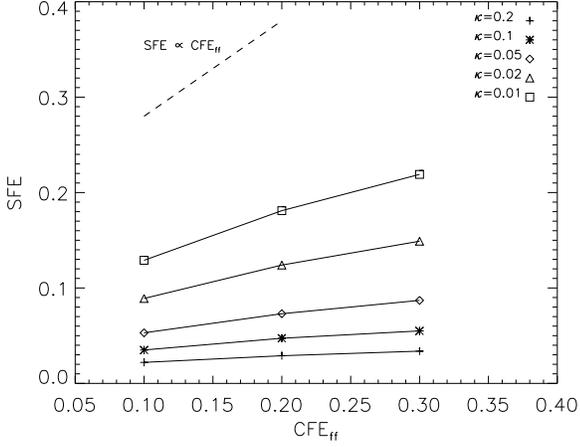,width=\columnwidth} 
\end{center}
\caption{Dependence of the star formation efficiency SFE on the value of core formation efficiency per unit free-fall time, CFE$_{ff}$ in models with uniform star formation histories (i.e., constant CFE per unit time). The models are shown for different values of the wind efficiencies factor, $\kappa$. The dashed line in the upper left corner of the figure marks what would be a linear relationship between the SFE and the CFE$_{ff}$. The mass of the clump in all of these models is $10^{5}$ M$_{\odot}$ and the metallicity is $Z=Z_{\odot}$.}
\label{fig4}
\end{figure}

Fig.~\ref{fig1} displays the detailed time evolution of the CMF and of the IMF in one of the models. In this model, the clump mass is $M_{clump}=10^{5}$ M$_{\odot}$ and the CFE per unit time is constant and is given by CFE$_{ff}=0.2$. The metallicity of the gas in this model is solar. The left and right hand panels in Fig.~\ref{fig1} display the time evolution of the CMF and of the IMF, respectively. Initially, no stars are present in the clump. By $t \approx 2\times10^{4}$ yrs, a fraction of the early populations of cores have already turned into stars (i.e., after their ages became larger than their local contraction timescale, $t_{cont,p} (r,M)$). With the formation of newer generation of stars,  the instantaneous SFE increases (i.e., Fig.~\ref{fig2}, dashed line) as well as the ratio of $E_{k,wind}/E_{grav}$ (dashed line in right panel of Fig.~\ref{fig3}). The expulsion of the gas from the protocluster region and the termination of the core and star formation processes occur whenever the ratio $E_{k,wind}/E_{grav}$ reaches unity. In Fig.~\ref{fig2}, the various symbols mark the position of the final value of the SFE for various values of $\kappa$ in the range $[0.01-0.2]$, where $\kappa$ is the wind efficiency parameter. This range of values for $\kappa$ is motivated by the fact that larger values lead to extremely small value of the SFE $\lesssim 1\%$, whereas smaller values of $\kappa$ lead to SFE values near unity. For $\kappa=0.01$, the expulsion of the gas occurs at $t \approx 4.4\times 10^{5}$ yrs, which corresponds to the last timestep shown in Fig.~\ref{fig1}. 

\begin{figure}
\begin{center}
\epsfig{figure=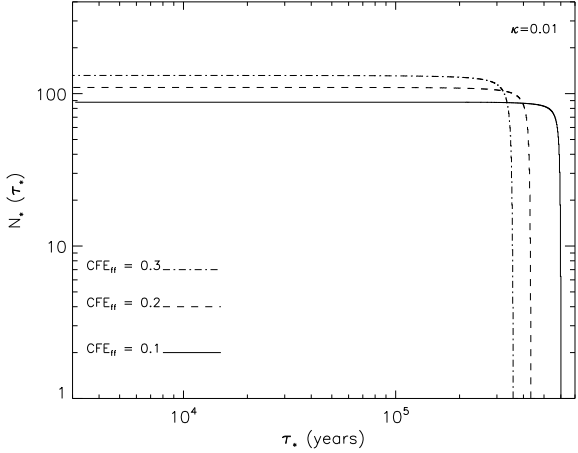,width=\columnwidth} 
\end{center}
\caption{Distributions of stellar ages in the uniform star formation models with various values of the CFE$_{ff}$ and for a value of the wind efficiency parameter, $\kappa=0.01$. The mass of the clump in all of these models is $10^{5}$ M$_{\odot}$ and the metallicity is $Z=Z_{\odot}$.}
\label{fig5}
\end{figure}

\begin{figure}
\begin{center}
\epsfig{figure=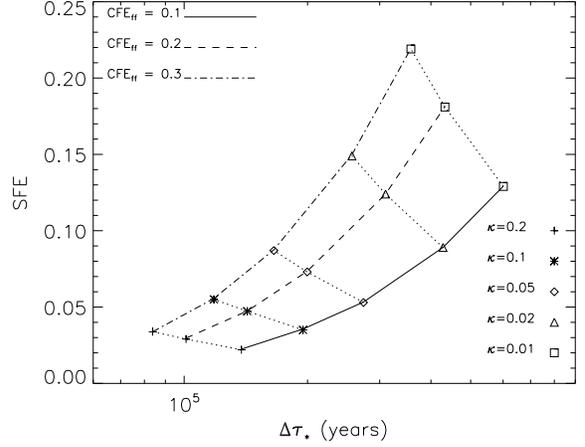,width=\columnwidth} 
\end{center}
\caption{SFE-$\Delta \tau_{*}$ diagrams for the uniform star formation models with various values of the CFE$_{ff}$. The different symbols refer to various values of the wind efficiency parameter. The mass of the clump in all of these models is $10^{5}$ M$_{\odot}$ and the metallicity is $Z=Z_{\odot}$.}
\label{fig6}
\end{figure}

\begin{figure}
\begin{center}
\epsfig{figure=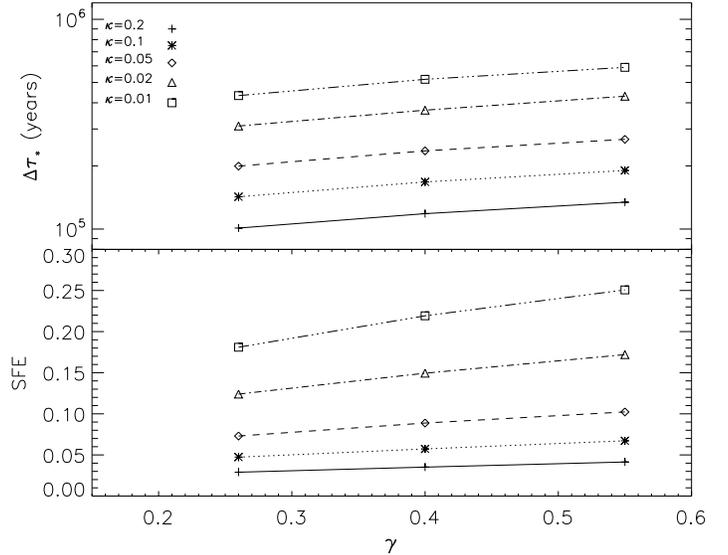,width=\columnwidth} 
\end{center}
\vspace{0.3cm}
\caption{Dependence of the SFE and age spread ($\Delta\tau_{*}$) on the turbulence parameter $\gamma$ shown for various choices of the wind efficiency parameter $\kappa$. The results shown here are for a model with a constant core formation efficiency per unit time of CFE$_{ff}=0.2$, a clump mass of $10^{5}$ M$_{\odot}$, and a metallicity of $Z=Z_{\odot}$.}
\label{fig7}
\end{figure}

Fig.~\ref{fig2} and Fig.~\ref{fig3} (right panel) also display the effect of changing the value of the CFE$_{ff}$ in the C-class models (all models are for a clump mass of $10^{5}$ M$_{\odot}$ and at solar metallicity). The figures show that higher values of the CFE$_{ff}$ lead to the earlier expulsion of the gas as feedback from stellar winds has a strong supra-linear dependence on stellar mass ($\propto M_{*}^{4}$, see Dib et al. 2011) and to a final SFE that does not scale linearly with the CFE$_{ff}$ (Fig.~\ref{fig4}). The choice of constant CFEs per unit time leads to flat distributions of stellar ages which are displayed in Fig.~\ref{fig5} (shown for $\kappa=0.01$). The lower numbers of stars observed for the older populations is simply due to the fact that cores have mass dependent, finite lifetimes of a few $10^{4}$ yrs. Fig.~\ref{fig6} displays the dual constraints obtained on the SFE and $\Delta \tau_{*}$ in the C-class models. The SFE-$\Delta \tau_{*}$ diagrams in Fig.~\ref{fig6} show an SFE that varies in the range [$0.025-0.25$] and age spreads that fall in the range $ [10^{5}-7\times 10^{5}]$ yrs. We explore the effect of changing the value of the turbulence parameter, $\gamma$, for which we have adopted a fiducial value of 0.26 (Kritsuk et al. 2007). Federrath et al. (2012) found a value of $\approx 0.33$ in numerical simulations of molecular clouds when turbulence is driven solely by solenoidal motions increasing up to a value of $\approx 1$ when turbulence is driven exclusively by compressive motions. Observationally, Brunt (2010) and Kainulainen (2013) derived values of $\gamma \approx 0.3-0.5$ in a number of nearby star forming clouds. We have performed a few additional models with values of $\gamma$ in the range [0.26-0.55] (while fixing the CFE$_{ff}=0.2$). The dual constraints on SFE-$\Delta \tau_{*}$ obtained for these various values of $\gamma$ are displayed in Fig.~\ref{fig7}. The effect of changing the value of $\gamma$ on the SFE and $\Delta \tau_{*}$ is modest and varies from a few percent to a few tens of percent depending on the adopted wind efficiency parameter $\kappa$. While the values of $\Delta \tau_{*}$ obtained above bracket the range of age spreads that are measured for massive starburst clusters (e.g., Kudryavtseva et al. 2012), the measured values of the SFE for these clusters are found to be larger (see \S.~\ref{compobs} and Tab.~\ref{tab2}). We therefore explore additional models in which the CFE increases with time. We consider the cases of a power law increase of the CFE with time (PL-models):

\begin{equation}
{\rm CFE} \left(t\right)=A\times\left( \frac{t}{t_{ff,cl}}\right)^{\alpha}, 
\label{eq13}
\end{equation}

\noindent and others where the CFE increases exponentially with time (EXP-models):

\begin{equation}
{\rm CFE} \left(t\right)=B\times{\rm exp}\left(\frac{1}{\beta} \frac{t}{t_{ff,cl}}\right).
\label{eq13}
\end{equation}

\begin{figure}
\begin{center}
\epsfig{figure=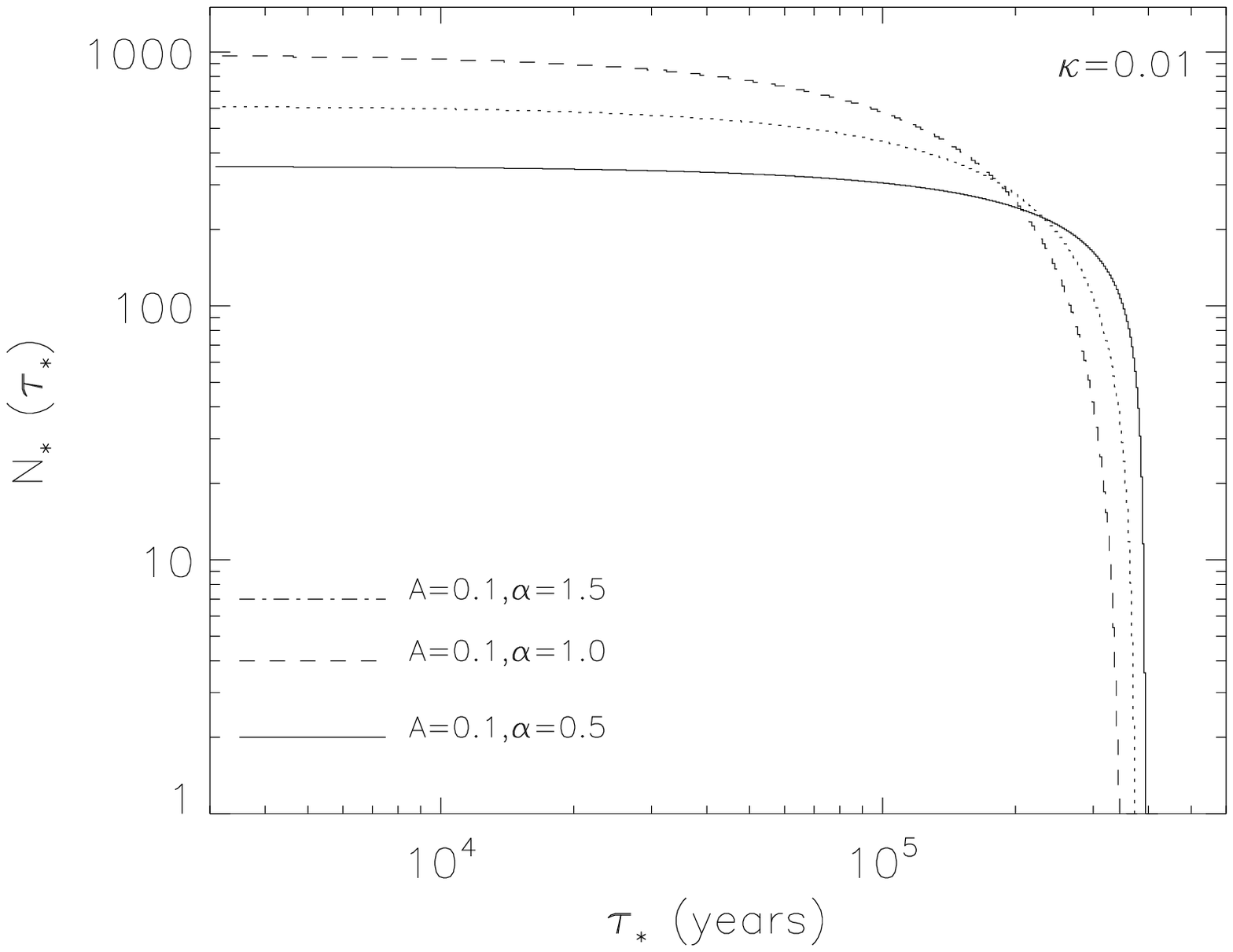,width=\columnwidth} 
\end{center}
\caption{Distributions of stellar ages in models with an increasing core formation efficiency per unit time (PL models) and for a value of the wind efficiency parameter, $\kappa=0.01$. The different lines refer to three permutations of the parameters that describe the power law shape of the CFE (Eq.~\ref{eq13}). The mass of the clump in all of these models is $10^{5}$ M$_{\odot}$ and the metallicity os $Z=Z_{\odot}$.}
\label{fig8}
\end{figure}

\begin{figure}
\begin{center}
\epsfig{figure=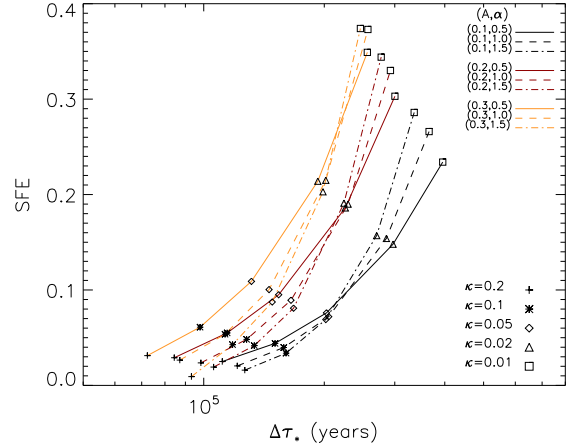,width=\columnwidth} 
\end{center}
\caption{SFE$-\Delta \tau_{*}$ diagrams for models with an increasing core formation efficiency per unit time. The different lines refer to a number of permutations of the parameters that describe the power law shape of the CFE (Eq.~\ref{eq13}). The various symbols refer to different values of the wind efficiency parameter, $\kappa$. The mass of the clump in all of these models is $10^{5}$ M$_{\odot}$ and the metallicity is $Z=Z_{\odot}$.}
\label{fig9}
\end{figure}

\begin{figure}
\begin{center}
\epsfig{figure=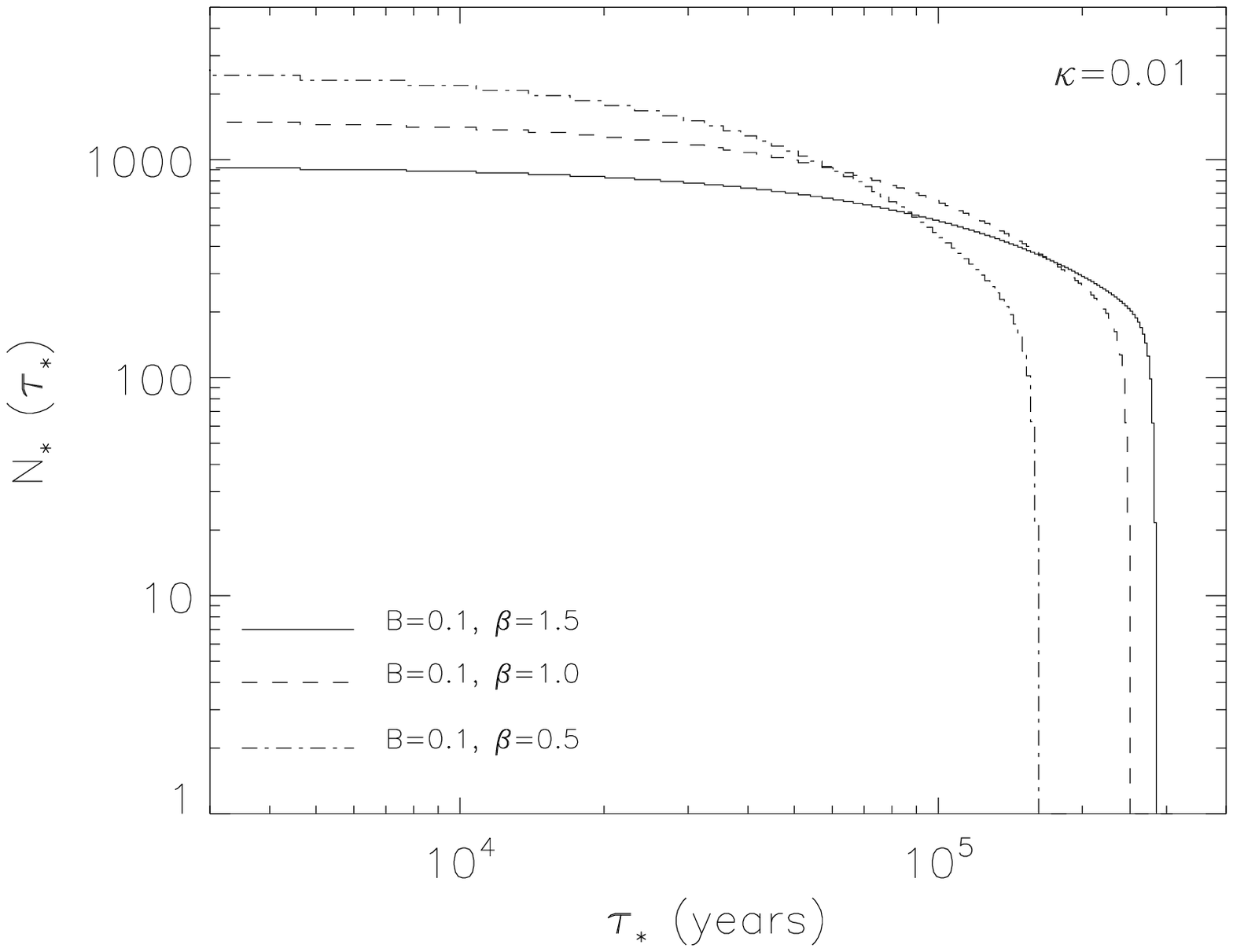,width=\columnwidth} 
\end{center}
\caption{Distributions of stellar ages in models with an increasing core formation efficiency per unit time (EXP models) and for a value of the wind efficiency parameter, $\kappa=0.01$. The different lines refer to three permutations of the parameters that describe the exponential shape of the CFE (Eq.~\ref{eq14}). The mass of the clump in all of these models is $10^{5}$ M$_{\odot}$ and the metallicity is $Z=Z_{\odot}$.}
\label{fig10}
\end{figure}

\begin{figure}
\begin{center}
\epsfig{figure=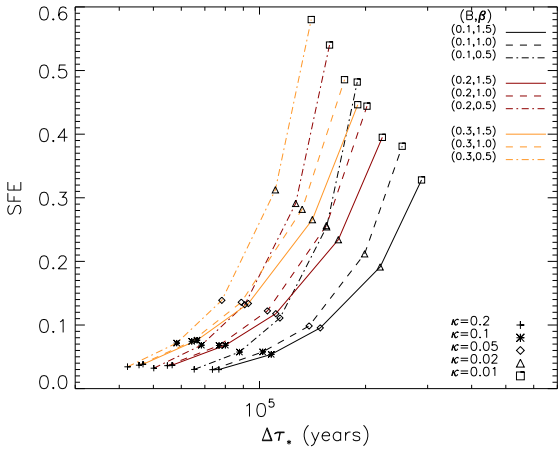,width=\columnwidth} 
\end{center}
\caption{SFE-$\Delta \tau_{*}$ diagrams for models with an increasing core formation efficiency per unit time (EXP models). The different lines refer to a number of permutations of the parameters that describe the exponential shape of the CFE (Eq.~\ref{eq14}). The various symbols refer to different values of the wind efficiency parameter, $\kappa$. The mass of the clump in all of these models is $10^{5}$ M$_{\odot}$.}
\label{fig11}
\end{figure}

Fig.~\ref{fig8} and Fig.~\ref{fig10} display the distributions of stellar ages for models with $M_{clump}=10^{5}$ M$_{\odot}$, $\kappa=0.01$, and for a few permutations of the parameters $(A-\alpha)$ and $(B-\beta)$. The SFE-$\Delta \tau_{\star}$ diagrams for the PL- and EXP-class models are displayed in Fig.~\ref{fig9} and Fig.~\ref{fig11}, respectively. These figures show that models with an accelerated mode of star formation tend to yield higher values of the SFE as well as smaller age spreads ($\Delta \tau_{*} \lesssim 3-4 \times 10^{5}$ ) as compared to the cases with a constant CFE per unit time, for any given value of $\kappa$. 

\subsection{INFLUENCE OF THE CLUMP MASS AND COMPARISON TO OBSERVATIONS}\label{compobs}

\begin{table*}
\begin{minipage}{\textwidth}
\begin{center}
\caption{Name of the clusters followed by their: age, age spread, half mass radius, photometric mass, measured 1D velocity dispersion, corrected 1D velocity dispersion (with a Kroupa IMF), corrected 1D velocity dispersion (with a shallow IMF: slope of -2.1 at the high mass end), dynamical mass (with a Kroupa IMF), dynamical mass (with a shallow IMF: slope of -2.1 at the high mass end), SFE (with a Kroupa IMF), SFE (with a shallow IMF: slope of -2.1 at the high mass end).}

\begin{tabular}{llllllllllll}
\hline
\hline
Cluster          & age                                                                                                    &	$\Delta \tau_{*}$                      &   r$_{\rm hm}$  &  M$_{\rm phot}$ & $\sigma_{\rm 1D,m}$ & $\sigma_{\rm 1D,c}$   & $\sigma_{1D,c}$   & M$_{\rm dyn}$ & M$_{\rm dyn}$  & SFE  &  SFE  \\ 
                       & [Myr]                                                                                                  & 	[Myr]                                           &   [pc]                   &  [$10^{3}$ M$_{\odot}$]   &  [km s$^{-1}]$               & [km s$^{-1}]$                & [km s$^{-1}$]          & [M$_\odot$]      & [M$_{\odot}$]    &           &            \\ 
                       &                                                                                                            &                                                             &                            &                            &                                        & Kroupa\footnotemark[13]         & SIMF \footnotemark[14]                           & Kroupa               &  SIMF    & Kroupa & SIMF\\    
 \hline
ONC              & 1.0 -- 2.5\footnotemark[1]                                                              &  $0.96\pm0.19$                                &  $0.8\pm0.1$\footnotemark[2] & $1.8\pm0.1$\footnotemark[2] & $2.5\pm0.5$\footnotemark[1] \footnotemark[3] & - & - & $8700$ & - & $0.20$ & -  \\
NGC 3603    & 1.0--2.0\footnotemark[4] \footnotemark[5]                                  & $\le 0.1\pm0.03$\footnotemark[5] &  $0.5\pm0.05$\footnotemark[6] \footnotemark[11] & $13\pm3$\footnotemark[10] & $4.5\pm0.8$\footnotemark[4] & 8.61 & 7.36 & 64632 & 47274 & 0.20 & 0.27   \\
Westerlund 1&  3.0--5.0\footnotemark[7] \footnotemark[12] & $\le 0.4\pm0.05$\footnotemark[5] &  $1.1\pm0.1$\footnotemark[7] \footnotemark[8] & $52\pm5.2$\footnotemark[7] &$4.2\pm1.1$\footnotemark[9] & $7.79$ & $6.65 $ & $116459$ & $84851$ & 0.44 & 0.61  \\
\hline 
\end{tabular}

\footnotemark[1]{Hillenbrand \& Carpenter 2000}
\footnotemark[2]{Hillenbrand \& Hartmann 1998}
\footnotemark[3]{Jones \& Walker 1988}
\footnotemark[4]{Rochau et al.\ 2010}
\footnotemark[5]{Kudryavtseva et al. 2012}
\footnotemark[6]{Stolte et al.\ 2004}
\footnotemark[7]{Brandner et al.\ 2008}
\footnotemark[8]{Gennaro et al.\ 2011}
\footnotemark[9]{Kudryavtseva 2012}
\footnotemark[10]{Stolte et al. 2006}
\footnotemark[11]{Harayama et al. 2008}
\footnotemark[12]{Negueruela et al. 2010}
\footnotemark[13]{values derived using a Kroupa IMF}
\footnotemark[14]{values derived using an IMF with a Shallower (SIMF) slope of -2.1 in the intermediate to high mass end.}

\label{tab2}
\end{center}
\end{minipage}
\end{table*}

\begin{figure*}
\begin{center}
\epsfig{figure=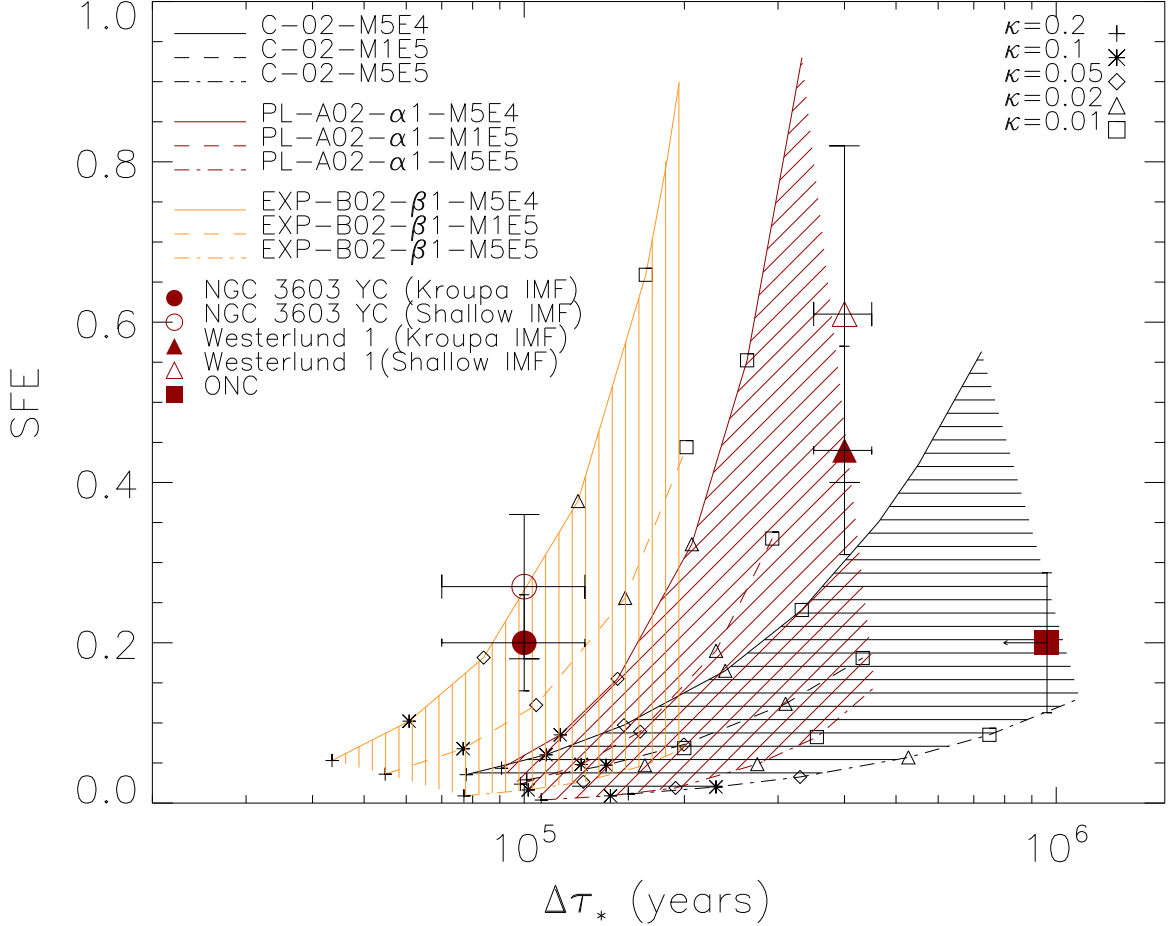,width=\textwidth} 
\end{center}
\caption{SFE-$\Delta \tau_{*}$ diagrams for models with different clumps masses of $5\times 10^{4}$ M$_{\odot}$ (full lines), $10^{5}$ M$_{\odot}$ (dashed lines), and $5\times10^{5}$ M$_{\odot}$ (dot-dashed line), and for different core and star formation histories in the clump (black: constant CFE per unit time, red: power law increasing CFE per unit time, and yellow exponentially increasing CFE per unit time). The various symbols refer to different values of the wind efficiency parameter, $\kappa$. The models are compared to the observational data of the starburst clusters NGC 3603 YC and Westerlund 1 and to the ONC. The metallicity in all of the models is $Z=Z_{\odot}$.} 
\label{fig12}
\end{figure*}

In all previous models, the mass of the protocluster clump was fixed to $10^{5}$ M$_{\odot}$. In this section, we investigate the effect of the mass of the clump on its position in the SFE-$\Delta \tau_{*}$ diagram. If we were to assume, as a crude approximation, that the clump has a uniform density (this is not the case in this work as the clumps have a radial density profile given by Eq.~\ref{eq1}), then its gravitational energy will be $E_{grav} \propto M_{clump}^{2}/R_{c} \propto M_{clump}^{\sim 1.60}$ using the scaling relation between the masses and sizes of the clumps ($M_{clump} \propto R_{c}^{2.54}$, see \S.~\ref{clusters}). The effective wind energy dependence on the mass of the clump is $E_{k,wind} \propto M_{*}^{4} =$ SFE$^{4} M_{clump}^{4}$. The ratio $E_{k,wind}/E_{grav}$ is thus $\propto$ SFE$^{4} M_{clump}^{2.4}$. As the requirement of gas expulsion is that $E_{k,wind}/E_{grav}=1$ and since it is independent of the clump mass, this implies that the SFE $\propto M_{clump}^{-0.60}$. Thus, we should expect a decreasing SFE with increasing clump mass. Fig.~\ref{fig12} displays the SFE-$\Delta \tau_{*}$ diagrams for clumps of various masses in the range $5\times 10^{4}-5\times 10^{5}$ M$_{\odot}$ and for the three different adopted prescriptions of the core formation history with their respective parameter permutations. The figure clearly shows that, irrespective of the choice of the core formation history, the SFE decreases with increasing clump mass for any given value of $\kappa$. While models with a constant CFE per unit time yield maximum values of the SFE $\approx 0.5$ with age spreads of up to $\Delta \tau_{*} \lesssim 1 \times 10^{6}$ yr (for values of $\kappa < 0.01$), the SFE in models with accelerated modes of the CFE and for the lower clumps masses approaches values as high as $\approx 0.8-0.9$ in association with shorter  age spreads of the order $\Delta \tau_{*} \lesssim 4 \times 10^{5}$ yrs. The labelling of the models in Fig.~\ref{fig12} is indicative of the choice of parameters. For example, model EXP-B02-$\beta$1-M5E5 refers to a model with an exponential function for the time dependent CFE, along with the parameters $B=0.2$, $\beta=1$ and a clump mass of $5\times 10^{5}$ M$_{\odot}$.       

\begin{figure}
\begin{center}
\epsfig{figure=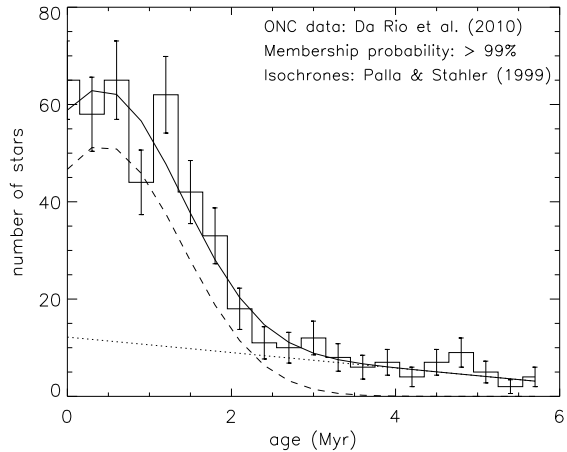,width=\columnwidth} 
\end{center}
\caption{Distribution of stellar ages in the Orion Nebula Cluster. The histogram is constructed using the data of Da Rio et al. (2010b) with a bin size in age of $3\times 10^{5}$ yrs. We considered only stars that have a cluster membership probability of 99 percent. The ages are determined using the  evolutionary tracks of Palla \& Stahler (1999). The data is simultaneously fitted with a function comprised of a linearly increasing star formation rate at old ages (dotted) and an accelerated (i.e., a burst) mode of star formation at young ages that is represented by a Gaussian function (dashed line). The global fit to the data is shown with the full line.}
\label{fig13}
\end{figure}

In Fig.~\ref{fig12}, we also compare our models to some of the available observational data, namely to the Orion Nebula Cluster (ONC) and to the more massive and dense clusters Westerlund 1 and NGC 3603 YC. Several properties of these three clusters are summarised in Tab.~\ref{tab2}. The age spreads for NGC 3603 YC and Westerlund 1 have been recently measured by Kudryavtseva et al. (2012). These authors used a Bayesian inference approach to constrain the age difference between the stellar populations in these clusters coupled to an assessment of the stars membership to the clusters from multi-epoch astrometric monitoring. They found age spreads of $\lesssim 0.1\pm 0.03$ Myrs and $\lesssim 0.4\pm 0.05$ Myrs for NGC 3601 YC and Westerlund 1, respectively. The determination of cluster membership and therefore of the true age spread in the ONC is more complex since the ONC stars move along the same direction as its background and foreground stars. The earlier work of Palla \& Stahler (2000) suggested that the $\Delta \tau_{*}$ in the ONC could be as large as $10$ Myrs (see however the criticism by Hartmann 2001;2003 and Preibisch 2012). We use the recent determinations of stellar ages for stars in the ONC of Da Rio et al. (2010b) derived by placing their positions on the Hertzsprung-Russel diagram with the Palla \& Stahler (1999) stellar evolutionary tracks. We construct the age distribution of stars in the ONC which is displayed in Fig.~\ref{fig13}. It is important to note that we do not have any direct way of assessing the contamination level in the ONC by foreground/background stars. However, we only consider in the construction of the age distribution of the cluster's stars those stars for which Da Rio et al. (2010b) have assigned a cluster membership probability of $\ge 99 \%$. We fit the stellar ages distribution in the ONC with a composite function comprising of a linearly increasing star formation rate at old ages and a Gaussian function for the bulk of the ONC stars at young ages (dotted and dashed lines in Fig.~\ref{fig13}, respectively). In this work, we consider that the width of the Gaussian function, which we find to be $0.96\pm0.19$ Myrs, is an appropriate representation of the age spread for the bulk of the stellar population in the ONC.    

Concerning the measurement of the SFE for these three clusters, the most obvious way would be to calculate the quantity $M_{*}/\left(M_{*}+M_{gas}\right)$, where $M_{*}$ is the measured photometric total stellar mass, and $M_{gas}$ the mass of the remnant gas in the clusters, respectively. However, in the case of the clusters considered here and in particular for the case of the older clusters such as Westerlund 1, the use of this equation to derive the SFE is not practical due to the absence of gas in the cluster. An alternative way of assessing the SFE is by assuming that the clusters have been re-virialized following the expulsion of the gas and with the additional assumption that the currently observed velocity dispersion of the stars is not very different from their value prior to the gas expulsion. If these assumptions are valid, then the measured dynamical mass could be an appropriate representation of the mass of the virialized system (stars+gas) at the stage where gas was still bound to the cluster. The dynamical mass is measured as being: 

\begin{equation}
M_{dyn}=\frac{\eta r_{hm} \sigma_{3D}^{2}}{G},
\label{eq15}
\end{equation}

\noindent where $\eta$ is a value that could depend on the stellar density radial profile, $r_{hm}$ is the half-mass radius of the cluster, $\sigma_{3D}$ is the three-dimensional velocity dispersion of the stars in the cluster, and $G$ is the gravitational constant. We adopt a value of $\eta \approx 2.5$ which was shown by Spitzer (1987) to depend only weakly on the cluster's radial stellar density profile. NGC 3606 YC is observed to be mass segregated and the core radius of the cluster increases when decreasing stellar masses are considered (N\"{u}rnberger \& Petr-Gotzens 2002). The value of $r_{hm}$ derived for the high mass stars from HST observations yield a value of $r_{hm} \approx 0.2$ pc (Stolte et al. 2004) while Harayama et al (2008) estimated a value of $r_{hm} \approx 0.7-1.5$ pc for stars in the mass range $0.5-2.5$ M$_{\odot}$ using near-infrared adaptive optics observations. In this work, we adopt an intermediate value of $r_{hm}=0.5\pm0.05$ pc. In the case of Westerlund 1, the half-mass radius of the cluster has been estimated by Brandner et al. (2008) and Gennaro et al. (2011) to be $r_{hm}=1.1 \pm 0.1$ pc. For the ONC, we adopt the value of $r_{hm}=0.8 \pm 0.1$ pc that has been derived by Hillenbrand \& Hartmann (1998).  

In this work, the adopted stellar velocity dispersions are ones that are measured from the stars proper motions rather than radial velocity measurements as the latter can be contaminated by orbital motions in binary systems and may tend to overestimate the measured velocity dispersions. In the ONC, Jones \& Walker (1988) found a mean velocity dispersion of $\left<2.34 \pm 0.09\right>$ km s$^{-1}$for $\sim 900$ stars brighter than $I \leq 16$ (i.e., down to $\approx 0.2$ M$_{\odot}$). Hillenbrand \& Hartmann (1998) assumed that the kinetic energy of the stars are in equipartition and used this to reassess the dependence of the stellar velocity dispersion on stellar mass using three distinct mass ranges. They found that $\left< \sigma_{1D}\right>=2.81$ km s$^{-1}$ in the mass range $0.1 < M/M_{\odot} < 0.3$ and $\left< \sigma_{1D}\right>=2.24$ km s$^{-1}$ over the mass range $1 < M/M_{\odot} < 3$. Here we adopt a value of $\sigma_{1D}=2.5 \pm 0.5$ km s$^{-1}$ for the entire stellar mass range in the ONC. The proper motions of stars in NGC 3603 YC and in Westerlund 1 have been measured for stars only in the limited mass ranges of [$1.7-9$] M$_{\odot}$ and $[1.5-10]$ M$_{\odot}$, respectively. In NGC 3603 YC, the measured velocity dispersions do not seem to depend on the stellar mass which may seem to indicate a lack of kinetic energy equipartition between the stars. However, the mass ranges over which these measurements have been performed in both clusters are not very large and at present it is not possible to rule out that equipartition may or not be present when extended to the entire stellar mass range. Keeping this in mind, we re-compute the velocity dispersions over the entire mass ranges. Assuming that the equipartition of kinetic energy between cluster members holds, this implies: 

\begin{equation}
\sigma_{1D,a}^{2}=\sigma_{1D,m}^{2} \frac{\left<M\right>_{m}} {\left<M\right>_{a}}, 
\label{eq16}
\end{equation}

\noindent where $\sigma_{1D,m}$ is the 1D velocity dispersion that has been measured over a given mass range and  $\sigma_{1D,a}$ is the 1D velocity dispersion of stars over any other mass range for which proper motions of stars have not been measured and $\left<M\right>_{m}$ and $\left<M\right>_{a}$ are the mean masses of stars over these mass ranges, respectively. For NGC 3603 YC, the missing mass ranges in the calculation of the velocity dispersion are [$M_{min}$-$1.7$ M$_{\odot}$] and [$9$ M$_{\odot}$-$M_{max}$] whereas for Westerlund 1 these ranges are [$M_{min}-1.5$ M$_{\odot}$] and [$10 $M$_{\odot}-M_{max}$]. For consistency with the calculation of the ONC, we make the assumption that the minimum mass in both clusters is also $M_{min}=0.2$ M$_{\odot}$. The maximum currently observed stellar mass in NGC 3603 YC, that of the star NGC 3603-A1, is $\approx 116\pm 31$ M$_{\odot}$ (Schnurr et al. 2008). The estimated mass loss rate of NGC 3603-A1 is $\approx 3.2 \times10^{-5}$ M$_{\odot}$ yr$^{-1}$ (Crowther et al. 2010). However, the mass loss rates of massive stars are known to increase by a factor of a few to several as they evolve on and off the main sequence (e.g., Voss et al. 2009), thus, we adopt a time averaged mass-loss rate of $10^{-5}$ M$_{\odot}$ yr$^{-1}$ over the lifetime of NGC 3603-A1. Considering the cluster's age of $1-2$ Myrs, this would imply that the initial mass of NGC 3603-A1 is $M_{max} \approx 125-135$ M$_{\odot}$, and therefore we adopt a mean value of 130 M$_{\odot}$. The maximum stellar mass in Westerlund 1 that  is currently observed is $\approx 27-30$ M$_{\odot}$ (Brandner et al. 2008) and may originate from a progenitor star with a mass of $\approx 30-40$ M$_{\odot}$. However, as the age of Westerlund 1 is $4-5$ Myrs, the most massive stars would have already evolved into the supernova phase. As Westerlund 1 is more massive than NGC 3603 YC, we therefore also assume that $M_{max}$ in Westerlund 1 is $\approx 130$ M$_{\odot}$. Using Eq.~\ref{eq16}, the corrected velocity dispersions computed for the entire stellar mass range can be calculated by assuming that: 

\begin{equation}
M_{cluster} \sigma_{1D,c}^{2} \approx M_{m} \sigma_{1D,m}^{2}+ \sum_{i=1}^{n} M_{a}(i) \sigma_{1D,a}(i)^{2},
\label{eq17}
\end{equation}   

\noindent where $M_{m}$, and $M_{a}(i)$ are the total stellar mass of the cluster in the mass ranges over which proper motions measurements are available and missing, respectively, and where the index $i$ refers to all missing mass ranges (here $n=2$). Eq.~\ref{eq17} can be simply re-written as being: 

\begin{eqnarray}
\sigma_{1D,c}^{2}=\frac{M_{m}}{M_{cluster}} \sigma_{1D,m}^{2}+\sum_{i=1}^{n} \frac{M_{a}(i)}{M_{cluster}} \sigma_{1D,a}(i)^{2}\nonumber \\ 
~~~~~~~~~~~~~~= \frac{N_{m} \left<M\right>_{m}}{N_{tot} \left<M\right>} \sigma_{1D,m}^{2}+\sum_{i=1}^{n}\frac{N_{a}(i) \left<M_{a}(i)\right>}{N_{tot} \left<M\right>} \sigma_{1D,a}(i)^{2}\nonumber \\ 
~~~~~~~~~~~~~~= f_{m} \frac{\left<M_{m}\right>}{\left<M\right>} \sigma_{1D,m}^{2}+ \sum_{i=1}^{n} f_{a}(i)\frac{\left<M_{a}(i)\right>}{\left<M\right>} \sigma_{1D,a}(i)^{2}.
\label{eq18}
\end{eqnarray}

In Eq.~\ref{eq18}, $f_{m}$ and $f_{a} (i)$ (for the mass range $i$) represent the number fractions of stars in the cluster for which proper motion have been/not been measured, respectively. These fractions, along with the average stellar masses over the different mass ranges can be easily calculated from the IMF, provided the IMF is fully sampled, which is very likely to be the case in the massive clusters that are considered here. Over the entire mass range of $[0.2-130]$ M$_{\odot}$, we assume that the IMF is well represented by a) a Kroupa IMF (Kroupa 2002, Weidner \& Kroupa 2004) given by:

\begin{eqnarray}
\begin{array}{l} 
\xi\left(M\right)=k\\
\end{array}
\left\{
\begin{array}{l}
\left(\frac{M} {M_{H}}\right)^{-\alpha_{1}}, ~~~~~~~~~~~~~~~~~~ M_{min} \leq M \leq M_{0} \\
\left(\frac{M_{0}}{M_{H}}\right)^{-\alpha_{1}} \left(\frac{M}{M_{0}}\right)^{-\alpha_{2}},~~~~~ M _{0} \leq M \leq M_{1}\\
\left(\frac{M_{0}}{M_{H}}\right)^{-\alpha_{1}} \left(\frac{M_{1}}{M_{0}}\right)^{-\alpha_{2}} \left(\frac{M}{M_{1}}\right)^{-\alpha_{3}}, M_{1} \leq M \leq M_{max} 
\end{array}
\right.  
\label{eq19}
\end{eqnarray}   

\noindent with $M_{H}=0.08$ M$_{\odot}$, M$_{0}=0.5$ M$_{\odot}$, M$_{1}=1$ M$_{\odot}$, and $\alpha_{1}=1.30$, $\alpha_{2}=2.30$, and $\alpha_{3}=2.35$ and b) a shallower than Salpeter IMF in the intermediate to high mass end. The derived IMFs for massive starburst clusters such as NGC 3603 YC, Westerlund 1, and the Arches cluster tend to indicate that the slopes of their IMF in the intermediate to high mass regime is potentially shallower than the Salpeter value (Stolte et al. 2006; Dib et al. 2007b, Dib 2007; Dib et al. 2008b, Pang et al. 2013). In our case, the functional form of this IMF is given by:

\begin{eqnarray}
\begin{array}{l} 
\xi\left(M\right)=k\\
\end{array}
\left\{
\begin{array}{l}
\left(\frac{M} {M_{H}}\right)^{-\alpha_{1}},~~~~~~~~~~~~~~M_{H} \leq M \leq M_{0} \\
\left(\frac{M_{0}}{M_{H}}\right)^{-\alpha_{1}}  \left(\frac{M}{M_{0}}\right)^{-\alpha_{sb}}, M _{0} \leq M \leq M_{max}
\end{array}
\right.  
\label{eq20}
\end{eqnarray}   

\noindent with $\alpha_{sb}=2.1$. For both IMFs described by Eq.~\ref{eq19} and Eq.~\ref{eq20}, we can calculate the relative numbers of stars in any given mass range [$M_{1}-M_{2}$] as being:

\begin{equation}
f_{\left(M_{1}-M_{2}\right)}=\frac{\int^{M_{2}}_{M_{1}} \xi \left(M\right) dM} {\int^{M_{max}}_{M_{min}} \xi \left(M\right) dM},
\label{eq21}
\end{equation} 
  
\noindent and the mean mass over the corresponding range of masses as being: 

\begin{equation}
\left<M\right>_{\left(M_{1}-M_{2}\right)}=\frac{\int^{M_{2}}_{M_{1}} M \xi \left(M\right) dM}  {\int^{M_{2}}_{M_{1}} \xi \left(M\right) dM}.
\label{eq22}
\end{equation}
      
Using Eqs.~\ref{eq16}-\ref{eq22} and the measured values of the velocity dispersions, $\sigma_{1D,m}$, over the mass ranges of [$1.7-9$] M$_{\odot}$ and [$1.5-10$] M$_{\odot}$ in NGC 3603 YC and Westerlund 1 (see Tab.~\ref{tab2}), we find that the corrected velocity dispersions are $8.61$ km s$^{-1}$ and $7.79$ km s$^{-1}$ for a Kroupa IMF and $7.36$ km s$^{-1}$ and $6.65$ km s$^{-1}$ for the shallower IMF, respectively. Using Eq.~\ref{eq15} and the half-mass radii of both clusters listed in Tab.~\ref{tab2}, we calculate dynamical masses of $64632$ M$_{\odot}$ and $116459$ M$_{\odot}$ with a Kroupa IMF and $47274$ M$_{\odot}$ and $84851$ M$_{\odot}$ for the shallower IMF, for NGC 3603 YC and Westerlund 1, respectively. With these measurements of the dynamical masses of the clusters, we calculate the SFE which is given by: 

\begin{equation}
{\rm SFE}  = \frac{M_{phot}} {M_{dyn}}, 
\label{eq23} 
\end{equation}
     
where $M_{phot}$ is the estimated mass of the clusters measured from photometry alone. The values of $M_{phot}$ for the three clusters considered here are listed in Tab.~\ref{tab2}. Applying Eq.~\ref{eq23} yields an SFE$=0.20$ for the ONC and $0.20$ and $0.44$ for NGC 3603 YC and Westerlund 1 with a Kroupa IMF and $0.27$ and $0.61$ with the shallower IMF for the two starburst clusters. The uncertainties on the SFE values are assumed to be associated with the combination of uncertainties on $M_{phot}$, $\sigma_{1D,m}$, and $r_{hm}$. Assuming that the errors on each of these latter quantities are uncorrelated, $\delta \left({\rm SFE} \right)$ is given by:

\begin{eqnarray}
\delta\left({\rm SFE}\right)=\left(\frac{G}{3 \eta r_{hm} \sigma_{1D}^{2}}\right) \times \nonumber \\
\left[\left(\delta M_{phot}\right)^{2}+\left(\frac{M_{phot}^{2}}{r_{hm}^{2}} \left(\delta r_{hm}\right)^{2}\right)+\left(\frac{4 M_{phot}^{2}}{\sigma_{1D}^{2}} \left(\delta \sigma_{1D}\right)^{2}\right) \right]^{1/2}.
\label{eq24}
\end{eqnarray}

The data points for the ONC, NGC 3603 YC and Westerlund 1 are placed on the SFE-$\Delta \tau_{*}$ diagram in Fig.~\ref{fig12} along with the error bars calculated using Eq.~\ref{eq24}. The positions of NGC 3603 YC and Westerlund 1 are very well reproduced by models with accelerated modes of star formation whereas the position of the ONC is well reproduced by models in which star formation in the protocluster clump is constant in time. The point for NGC 3603 YC falls between the clump models with masses in the range $[5\times 10^{4}-10^{5}]$ M$_{\odot}$ and with an exponential mode of the core formation efficiency. This position is well matched by its calculated dynamical mass, $M_{dyn}$ (i.e., an approximation of the original clump mass), which is in the range $[4.7-6.4]\times10^{4}$ M$_{\odot}$ (depending on the adopted IMF) and it also implies a value for the wind efficiency parameter in the range $\kappa \approx[0.02-0.05]$. A similar good agreement is also found for Westerlund 1 for which the estimated dynamical mass falls in the range $[0.84-1]\times10^{5}$ M$_{\odot}$ and which is well bracketed by the clump models with masses between $[10^{5}-5\times10^{5}]$ M$_{\odot}$. The implied value for the wind efficiency parameter is however smaller in Westerlund 1 (i.e., $k\approx 0.005$). The agreement between the models and the ONC is more problematic as the point falls between clump models with masses between $[10^{5}-5\times10^{5}]$ M$_{\odot}$ while its estimated dynamical mass is only $8.7\times10^{3}$ M$_{\odot}$. The agreement for the ONC can be improved if the true age spread in the ONC was smaller than its currently measured value as the ONC membership is susceptible of being affected by contamination by older foreground stars (e.g., F\H{u}r\'{e}sz et al. 2008). Alternatively, the agreement can also be improved if the SFE in the ONC is higher than the value derived from the dynamics of stars alone and if the SFE may continue to grow as gas continues to fall towards the cluster centre (e.g., F\H{u}r\'{e}sz et al. 2008). We should point out that in all models displayed in Fig.~\ref{fig12}, we have employed the fiducial normalisation value for the CFE, either CFE$_{ff}=0.2$ in models with a constant core formation per unit time or $A=0.2$ and $B=0.2$ in models where the CFE is time dependent. The variations of this quantity by a factor of $2-3$ can cause a shift by a factor of a few in both the SFE and $\Delta \tau_{*}$ as already shown in Figs.~\ref{fig6}, \ref{fig9}, and \ref{fig11} for a given value of the wind efficiency parameter, $\kappa$.   
      
 \subsection{INFLUENCE OF METALLICITY: PREDICTIONS}

\begin{figure}
\begin{center}
\epsfig{figure=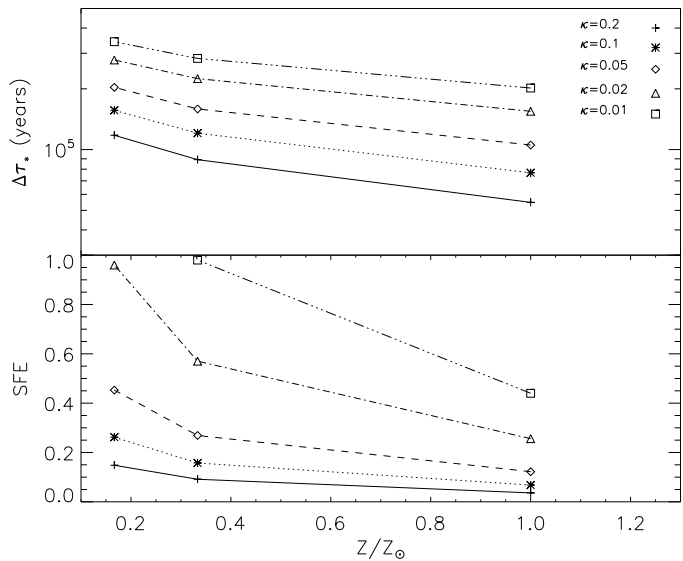,width=\columnwidth} 
\end{center}
\vspace{1cm}
\caption{Dependence of the stellar age spread (top), and SFE (bottom) on the metallicity of the clump in which stars form.  In this model the mass of the clump is $10^{5}$ M$_{\odot}$ and the core formation efficiency increases exponentially as a function of time (i.e., Eq~\ref{eq14}) with $B=0.2$ and $\beta=1$.}
\label{fig14}
\end{figure}

As already pointed out in Dib et al. (2011), the winds of massive stars are strongly metallicity dependent (Vink et al. 2001; Bresolin \& Kudritzki 2004) with the wind power that scales nearly linearly with metallicity (see Fig.~\ref{fig5} in Dib et al. 2011). The weaker power of the winds at lower metallicities, and the resulting longer expulsion timescales of the gas from the protocluster forming region, will cause the formation of additional generations of stars and lead to a higher final SFE. Thus, it is expected that the position of clusters formed in low metallicity environments in the SFE-$\Delta \tau_{*}$ diagram will be shifted towards both higher SFE and higher age spreads. We have performed two additional calculations for protocluster forming regions with the metallicities of $1/3$ and $1/6$ the solar metallicity value and which are typical of those found in the Large and Small Magellanic Clouds, respectively. The wind powers as function of stellar mass for these additional metallicity values are calculated using the fourth order polynomial analytical fits provided in Dib et al. (2011). Both models have a clump mass of $10^{5}$ M$_{\odot}$ and a time dependent CFE that increases exponentially with time (with $B=0.2$, and $\beta=1$) and are compared to the model with the corresponding solar metallicity model. Fig.~\ref{fig14} displays the variations of the SFE and $\Delta \tau_{*}$ as a function of metallicity. The predicted SFE and $\Delta \tau_{*}$ values show a clear dependence on metallicity and increase with decreasing metallicity. Future observations of a larger number of resolved massive extragalactic clusters should be able to test this prediction.    
  
\section{CONCLUSIONS AND DISCUSSION}\label{summary}

In this work, we investigate how stellar feedback in the form of stellar winds from massive stars imposes dual constrains on the star formation efficiency (SFE) and the age spread of stars ($\Delta \tau_{*}$) in massive clusters. We follow the co-evolution of the mass function of gravitationally bound cores and of the mass function of stars that form in a protocluster clump (IMF). We make the assumption that populations of gravitationally bound cores form at different epochs and at different positions from the protocluster clump centre by gravo-turbulent fragmentation. The cores collapse to form stars after evolving over their lifetimes which are taken to be a few times their free-fall times. The population of newly formed OB stars power strong stellar winds that inject significant amounts of kinetic energy in the clump. A fraction of this energy will counter the gravitational binding energy of the clump and act to disperse gas from the protocluster clump and hence quench further core and star formation. The expulsion of the gas from the protocluster clump sets the final SFE and the value of $\Delta\tau_{*}$. We consider the effect of varying the core formation efficiency per unit time going from models with a constant CFE per unit time to models in which the CFE increases with time following parametrized power law and exponential functions. Our results suggest that models with a constant CFE per unit time result in moderate values of the SFE (i.e., up to 0.3) and age spreads that are $\Delta \tau_{*} \lesssim 1$ Myrs. In contrast, models with a accelerated mode of core (and star) formation can lead to values of the SFE as high as $\approx 0.8-0.9$ with typically smaller age spreads $\Delta \tau_{*} \lesssim 0.4$ Myrs. We show that the final SFE in the clusters depends on the efficiency of the feedback by the winds and that it increases sub-linearly with an increasing CFE per unit time. We also show that the SFE decreases with the increasing mass of the clump. Both effects are due to the supra-linear dependence of the energy input rate by winds on stellar mass $\dot{E}_{wind} \propto M_{\star}^{4}$.  

We compare (i.e., in Fig.~\ref{fig12}) the constraints on the SFE and $\Delta \tau_{\star}$ of the multi-clump mass, multi-wind efficiency, and multi-core formation histories models to some of the available observational data for massive clusters in which massive stars are suspected of playing a major role in the expulsion of the gas from the cluster region. The models are compared to the Orion Nebula Cluster (ONC) and to the more massive starburst clusters NGC 3603 YC and Westerlund 1. The positions of NGC 3603 YC and Westerlund 1 on the SFE-$\Delta \tau_{*}$ diagram are well reproduced by models in which the core formation efficiency in the clump increases strongly with time with an implied wind efficiency parameter (i.e., the fraction of the wind kinetic energy that opposes gravity) of [0.02-0.05] and of $< 0.01$ in the case of NGC 3603 YC and Westerlund 1, respectively. The position of the ONC in the SFE-$\Delta \tau_{*}$ diagram is better reproduced by models in which the CFE per unit time is constant. This difference may be indicative of different modes of star cluster formation. Starburst clusters such as Westerlund 1, NGC 3603 YC, and the Arches cluster possess stellar surface densities ($M_{phot}/(\pi r_{hm}^{2})$) in excess of $10^{4}$ M$_{\odot}$ pc$^{-2}$ ($\approx 1.6\times10^{4}$ M$_{\odot}$ pc$^{-2}$ for NGC 3603 YC and $\approx 1.3\times10^{4}$ M$_{\odot}$ pc$^{-2}$). The high stellar surface density requires bringing a large fraction of the gas of the clump beyond the critical point for gravitational collapse. This violent compression of a large fraction of the cloud/clump mass may occur as the result of cloud-cloud collisions in the galactic disk. The observations of Furuka et al. (2009) and more recently of Fukui et al. (2013) are strongly suggestive of the occurrence of this scenario. In contrast, the formation of the parental clump/cloud of an ONC- like cluster (with stellar surface densities of $\approx 8 \times 10^{2}$ M$_{\odot}$ pc$^{-2}$) may originate from a more generic large scale gravitational instability in the disk or from the formation of the clump/cloud by converging flows in the more diffuse interstellar medium, or from swept up material by the expanding shock from a supernova explosion. Simulations of clump/cloud formation in the converging flows scenario show indications that a small number of stars may form in the early phases  when the converging flows meet and before a self-gravitating molecular cloud is fully assembled (e.g., V\'{a}zquez-Semadeni et al. 2007). This early generation of stars falls in the deepening gravitational potential of the molecular cloud during its formation and where the bulk of the star formation will occur. This is a plausible scenario which can explain the larger age spreads observed in the ONC as compared to the case of the starburst clusters. Since many star formation models are able to reproduce one or the other of the observational products of the star formation process, we suggest that placing a stellar cluster on the SFE-$\Delta \tau_{*}$ diagram is a powerful method to distinguish the dominant mode of star cluster formation in different environments and a very useful tool for testing star formation theories.    
   
 \section*{Acknowledgments}

We would like to thank the Referee for her/his constructive comments and useful suggestions. 

{}

\label{lastpage}

\end{document}